\documentclass[sigconf,natbib=true,anonymous=false]{acmart}
\usepackage{extarrows}
\usepackage{color} 
\usepackage{xcolor}
\usepackage{makecell}
\usepackage{amsmath,amsfonts}
\usepackage{algorithmic}
\usepackage{array}
\usepackage{graphicx}
\usepackage{stfloats}
\usepackage{multicol}
\usepackage{multirow}
\usepackage{hhline}
\usepackage{colortbl}
\usepackage{bm}
\usepackage{url}
\usepackage{enumitem}
\usepackage{arydshln}
\usepackage{bbding}
\usepackage{utfsym}
\usepackage[most]{tcolorbox}

\definecolor{lightgray}{rgb}{0.88, 0.92, 0.98}
\definecolor{defblue}{rgb}{0.1843, 0.3333, 0.6}
\definecolor{defred}{rgb}{0.88, 0.2510, 0.3294}

\AtBeginDocument{%
  \providecommand\BibTeX{{%
    \normalfont B\kern-0.5em{\scshape i\kern-0.25em b}\kern-0.8em\TeX}}}

\copyrightyear{2024}
\acmYear{2024}
\setcopyright{acmlicensed}\acmConference[SIGIR '24]{Proceedings of the 47th International ACM SIGIR Conference on Research and Development in Information Retrieval}{July 14--18, 2024}{Washington, DC, USA}
\acmBooktitle{Proceedings of the 47th International ACM SIGIR Conference on Research and Development in Information Retrieval (SIGIR '24), July 14--18, 2024, Washington, DC, USA}
\acmDOI{10.1145/3626772.3657727}
\acmISBN{979-8-4007-0431-4/24/07}

\begin{document}

\title{Simple but Effective Raw-Data Level Multimodal Fusion for Composed Image Retrieval}

\author{Haokun Wen}
\orcid{0000-0003-0633-3722}
\affiliation{
  \institution{\small School of Computer Science and Technology, Harbin Institute of Technology (Shenzhen)}
  \city{Shenzhen}
  \country{China}
  }
\email{whenhaokun@gmail.com}

\author{Xuemeng Song}
\orcid{0000-0002-5274-4197}
\authornote{Corresponding authors: Xuemeng Song and Liqiang Nie.}
\affiliation{%
  \institution{\small School of Computer Science and Technology, Shandong University}
  \city{Qingdao}
  \country{China}
  }
\email{sxmustc@gmail.com}

\author{Xiaolin Chen}
\orcid{0000-0003-4638-0603}
\affiliation{
  \institution{\small School of Software, Joint SDU-NTU Centre for \mbox{Artificial Intelligence Research, Shandong University}}
  \city{Jinan}
  \country{China}
  }
\email{cxlicd@gmail.com}

\author{Yinwei Wei}
\orcid{0000-0003-1791-3159}
\affiliation{%
  \institution{\small Faculty of Information Technology, \mbox{Monash University}}
  \city{Melbourne}
  \country{Australia}
  }
\email{weiyinwei@hotmail.com}

\author{Liqiang Nie}
\orcid{0000-0003-1476-0273}
\authornotemark[1]
\affiliation{%
  \institution{\small School of Computer Science and Technology, Harbin Institute of Technology (Shenzhen)}
  \city{Shenzhen}
  \country{China}
  }
\email{nieliqiang@gmail.com}

\author{Tat-Seng Chua}
\orcid{0000-0001-6097-7807}
\affiliation{%
  \institution{\small School of Computing, \mbox{National University of Singapore}}
  \country{Singapore}
  }
\email{dcscts@nus.edu.sg}

\begin{abstract}

Composed image retrieval (CIR) aims to retrieve the target image based on a multimodal query, \textit{i.e.}, a reference image paired with corresponding modification text.  Recent CIR studies leverage vision-language pre-trained (VLP) methods as the feature extraction backbone, and perform nonlinear feature-level multimodal query fusion to retrieve the target image. Despite the promising performance, we argue that their nonlinear feature-level multimodal fusion may lead to the fused feature deviating from the original embedding space, potentially hurting the retrieval performance. To address this issue, in this work, we propose shifting the multimodal fusion from the feature level to the raw-data level to fully exploit the VLP model's multimodal encoding and cross-modal alignment abilities. In particular, we introduce a Dual Query Unification-based Composed Image Retrieval framework (DQU-CIR), whose backbone simply involves a VLP model's image encoder and a text encoder. Specifically, DQU-CIR first employs two training-free query unification components: text-oriented query unification and vision-oriented query unification, to derive a unified textual and visual query based on the raw data of the multimodal query, respectively. The unified textual query is derived by concatenating the modification text with the extracted reference image's textual description, while the unified visual query is created by writing the key modification words onto the reference image. Ultimately, to address diverse search intentions, DQU-CIR linearly combines the features of the two unified queries encoded by the VLP model to retrieve the target image. Extensive experiments on four real-world datasets validate the effectiveness of our proposed method.
  
\end{abstract}



\maketitle

\section{Introduction}
\begin{figure}[htp]
        \vspace{-0.6em}
	\includegraphics[width=0.9\linewidth]{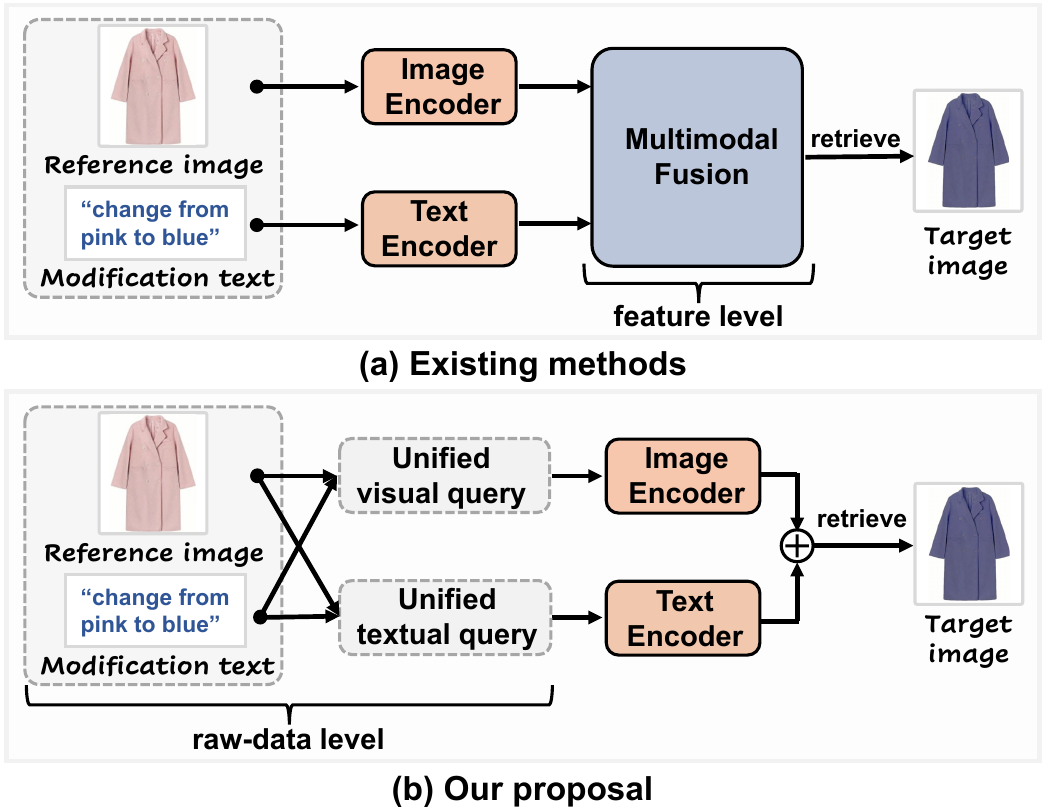}
        \vspace{-0.8em}
	\caption{Comparison between existing methods and ours.}
	\label{fig:pipline}
        \vspace{-1em}
\end{figure}

Different from traditional image retrieval that only allows users to express search intentions through a pure text or image query~\cite{lg1,lg2,yangxun1,yangxun2,liumeng}, composed image retrieval (CIR) enables users to utilize a multimodal query, \textit{i.e.}, a reference image with a modification text expressing some modification demands, to retrieve the target image, as exemplified in Figure~\ref{fig:pipline}. 
Given its promising application potential in many real-world scenarios, including intelligent robots~\cite{robot2} and commercial platforms~\cite{commer1, wl}, CIR has gained increasing research attention in recent years. The pipeline of the existing methods can be generally summarized in Figure~\ref{fig:pipline}(a). The reference image and the modification text are first processed by the image and text encoders, respectively. Thereafter, the extracted image and text features are fused through elaborately designed multimodal fusion functions to comprehend the users' search demands. Finally, the fused output is utilized to retrieve the target image. 

According to the type of utilized feature extraction backbone, existing efforts can be broadly classified into two groups: traditional model-based methods and vision-language pre-trained (VLP) model-based methods. The former utilizes traditional models like ResNet~\cite{resnet} or LSTM~\cite{lstm} to extract features for the reference/target image and the modification text, while the latter adopts the VLP model, especially those with dual encoders, such as CLIP~\cite{clip} and BLIP~\cite{blip}, as the feature extraction backbone. In contrast to traditional models, which are trained on single-modality datasets with limited scale, VLP models are pre-trained with large-scale image-text corpora, thereby achieving a more powerful capability in encoding both image and text information. 
Previous studies have consistently demonstrated the superiority of VLP models over the traditional feature extraction backbones in the CIR task~\cite{clip4cir,Liu_2024_WACV,tgcir}.

\begin{figure}[!t]
        \vspace{-0.5em}
	\includegraphics[width=0.87\linewidth]{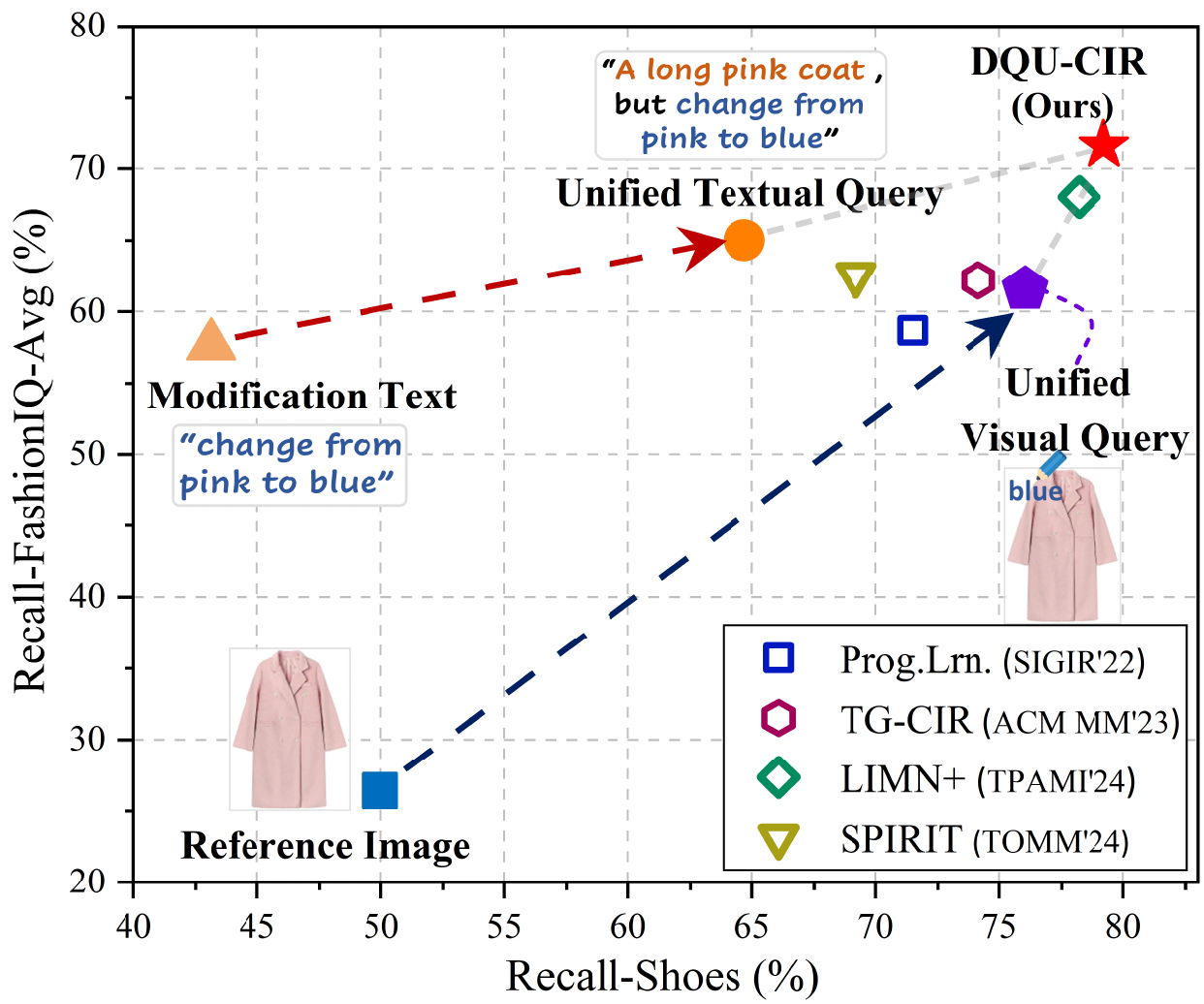}
        \vspace{-1.2em}
	\caption{Performance comparison of our method with state-of-the-art baseline methods on two public datasets.}
	\label{fig:intro_compare_baseline}
        \vspace{-1.4em}
\end{figure}

Although recent CIR methods have achieved promising performance with the advances in VLP models, they overlook the following crucial point. That is, benefiting from the image-text contrastive learning pre-training task, VLP models typically map the image and text into a common embedding space with corresponding encoders. It is apparent that this property of VLP models has naturally facilitated the multimodal query fusion and the subsequent target image retrieval of the CIR task, since the reference image, modification text, and target image can be projected into the common embedding space with VLP models' encoders. Nevertheless, existing VLP model-based CIR methods further employ nonlinear multimodal fusion functions to fuse the extracted image/text features, which can potentially cause  the fused multimodal query feature to deviate from the original common embedding space, thus hindering the target image retrieval.

Therefore, in this work, we propose to move the multimodal query fusion process from the feature level to the raw-data level,  to harness the full potential of VLP models. The main idea is to convert the multimodal query to a unified single-modal query that encapsulates the user's original search intention. This unified single-modal query can then be directly encoded by the VLP model's encoder and its feature can be used directly for 
target image retrieval, eliminating the need for feature-level multimodal query fusion. In this way, the multimodal encoding and cross-modal alignment abilities of the VLP model can be fully utilized. 

In particular, we design two query unification strategies: text-oriented unification and vision-oriented unification.  Text-oriented unification focuses on unifying the multimodal query into a pure text query, where any advanced image captioning model~\cite{blip2} can be used to convert the reference image into a high-quality textual caption and hence the unified textual query can be obtained by concatenating the generated reference image caption and the modification text. Vision-oriented unification targets composing the multimodal query into a unified image query by directly writing the target image description words mentioned in the modification text onto the reference image. In particular, we resort to the Large Language Model (LLM)~\cite{gemini} to capture the target image description words presented in the modification text, to take advantage of its outstanding natural language processing capability. Notably, the two query unification strategies are training-free, requiring no parameters and can be conducted offline.

To cope with the various search cases in real-world CIR tasks, we develop a dual query unification-based composed image retrieval framework (DQU-CIR), where the two unification strategies are jointly considered, as illustrated in Figure~\ref{fig:pipline}(b).
The underlying philosophy is that we expect that text-oriented query unification should be more useful for handling queries with complex search intentions, whereas  vision-oriented query unification excel at simpler search queries. This is because the user's complex search intention tends to be delivered by the modification text, while the unified textual query derived by the text-oriented query unification just contains the complete modification text, and should encapsulate the user's complex search intention better. In contrast, for simpler modification cases, the reference image typically plays a more significant role. Therefore, the output query of the vision-oriented unification, keeping the full content of the reference image, is more adept at capturing the user's search intention. 

Towards this end, DQU-CIR first conducts the text-oriented query unification and vision-oriented query unification in parallel. Subsequently, DQU-CIR encodes the unified textual and visual queries with the text and image encoder of the VLP model, respectively. Thereafter, DQU-CIR employs the linear weighted addition strategy to combine the features of the unified textual query and the unified visual query for target image retrieval. This linear fusion strategy is designed to ensure that the fused multimodal query embedding remains within the original embedding space of the VLP model, thus benefiting the final target image retrieval. To provide an intuitive demonstration of the effectiveness of our DQU-CIR, we illustrate the performance of our DQU-CIR and several state-of-the-art baseline methods on two public datasets (\textit{i.e.}, FashionIQ and Shoes) in Figure~\ref{fig:intro_compare_baseline}, where the performance of certain derivatives of our framework is also presented for better understanding the contribution of each key component. As can be seen, our DQU-CIR outperforms all the cutting-edge baselines even with a simple model design. Our main contribution can be summarized as follows.

\begin{itemize}[leftmargin=20pt]
	\item We design two training-free multimodal query unification strategies that can convert a multimodal query into a unified textual query and a unified visual query, respectively. To the best of our knowledge, we are the first to explore the raw-data level multimodal fusion in the context of CIR, which can fully leverage the VLP model's multimodal encoding and cross-modal retrieval capabilities. 

        \item  We propose a dual query unification-based composed image retrieval framework, named DQU-CIR, which effectively integrates the two query unification strategies for handling the real-world CIR cases with diverse search intentions. Notably, our framework consists of only a text encoder, an image encoder, and an MLP for learning the weights to linearly fuse the embeddings of the two unified queries.

        \item We conduct extensive experiments on four real-world public datasets, and the results demonstrate the superiority of our DQU-CIR over the state-of-the-art methods. In addition, we surprisingly found that directly writing descriptive words onto the image can achieve promising multimodal fusion results, which indicates the superior Optical Character Recognition (OCR) potential of the image encoder of the VLP model. We believe this would inspire the multimodal learning community to approach multimodal fusion from a new perspective.
        As a byproduct, we have released the source codes to benefit other researchers\footnote{\url{https://github.com/haokunwen/DQU-CIR}.}.
\end{itemize}

\section{Related Work}

\subsection{Vision-Language Pre-trained Models}

Inspired by the success of pre-trained models in the field of computer vision (CV) and natural language processing (NLP), many efforts have been dedicated to developing Vision-Language Pre-trained (VLP) models~\cite{vlp4,zx}. Benefiting from the pre-training on large-scale image-text corpora, VLP models yield universal cross-modal encoding capabilities, which have shown remarkable performance in various downstream multimodal tasks~\cite{vlp2, Liu_2024_WACV,vlp3}. Considering the inference speed for the retrieval task, in the context of CIR, we specifically focused on the VLP methods with the dual encoder structure~\cite{clip,align}. These models employ two single-modal encoders to encode images and text, respectively. 
Among them, the most representative method is CLIP~\cite{clip}, which is pre-trained on large-scale image-text pairs through the cross-modal contrastive learning task. It exhibits promising multimodal alignment and cross-modal retrieval capabilities, which are crucial for the CIR task. In this study, we have chosen CLIP as our feature extraction backbone.
In addition to its exceptional multimodal encoding capabilities, we have also observed notable natural language processing proficiency in its text encoder and Optical Character Recognition (OCR) potential in its image encoder. These properties of CLIP have enabled us to achieve promising CIR performance.

\subsection{Composed Image Retrieval}
According to the type of utilized feature extraction backbone, existing CIR methods can be classified into two groups: traditional model-based methods~\cite{tirg, val, cosmo, clvcnet,artemis, amc,comqueryformer,cmap} and VLP model-based methods~\cite{clip4cir,Liu_2024_WACV,tgcir,limn+,spirit,hq1,hq2}. The former employs the traditional models (e.g., ResNet and LSTM) to encode the given image and text. 
Due to the limited multimodal feature encoding capabilities of traditional models, methods of this group can only achieve suboptimal performance. While the latter group leverages the advanced VLP models for image/text encoding, which has shown more promising results due to their powerful multimodal feature encoding capabilities. A typical example is CLIP4CIR~\cite{clip4cir}, which first leverages CLIP to extract the image and text features, and then employs a combiner module to fuse the extracted features. Even with the simple design, CLIP4CIR also achieves satisfactory performance. Recently, Wen et al.~\cite{limn+} proposed LIMN+, which also employs CLIP as the feature extraction backbone. LIMN+ leveraged a masked Transformer~\cite{transformer} to fuse the extracted image/text features and based on that retrieve the target images. Additionally, it introduces a data augmentation technique to alleviate the issue of limited training samples are available for training the CIR models.

Although these methods have made prominent progress, they conduct multimodal fusion at the feature level with elaborate nonlinear functions, which may deviate the fused feature from the original embedding space. In contrast, we propose to achieve multimodal fusion at the raw-data level, which can be directly encoded by the advanced VLP methods. Then the superior multimodal encoding and cross-modal alignment capabilities of the VLP model can be fully utilized.

 \begin{figure*}[!t]
	\includegraphics[width=0.89\linewidth]{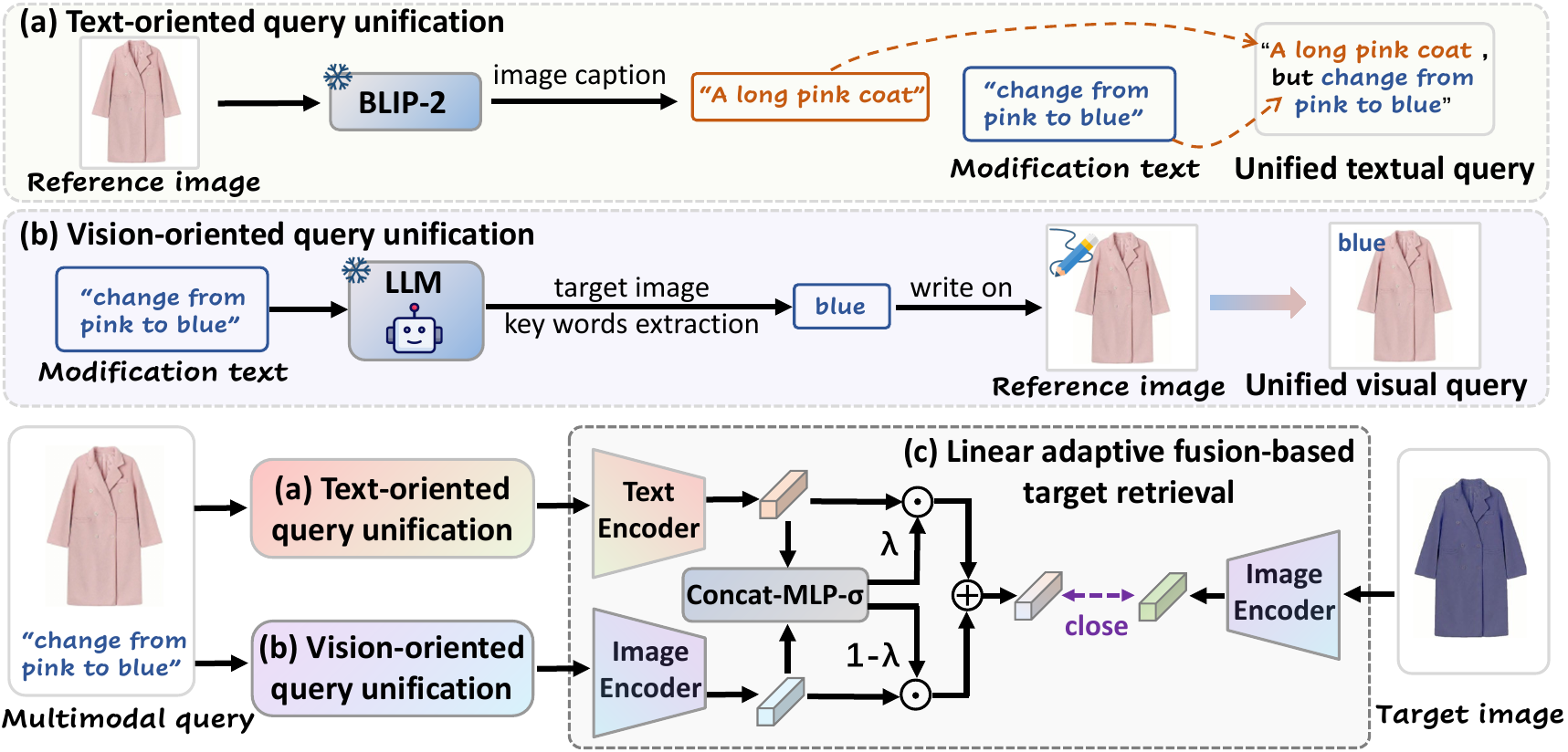}
        \vspace{-0.5em}
	\caption{The proposed DQU-CIR consists of three components: (a) text-oriented query unification, (b) vision-oriented query unification, and (c) linear adaptive fusion-based target retrieval.}
        \vspace{-0.8em}
	\label{fig:model}
\end{figure*}

\section{DQU-CIR}

In this section, we first formulate the problem and then detail DQU-CIR framework. As shown in Figure~\ref{fig:model}, DQU-CIR comprises three key components: text-oriented unification, vision-oriented unification, and linear adaptive fusion-based target retrieval.

\subsection{Problem Formulation}
In this work, we aim to address the task of CIR, which is to retrieve the target image that meets the multimodal query (\textit{i.e.}, a reference image, and a modification text). Formally, suppose we have a set of $N$ training triplets, denoted as $\mathcal{T}=\left\{ \left( {x_r}, {t_m}, {x_t} \right)_{i} \right\}_{i=1}^{N}$, where ${x_r}$, ${t_m}$, and ${x_t}$ refer to the reference image, modification text, and target image, respectively. Given a multimodal query $\left( {x_r}, {t_m} \right)$, our goal is to first derive a unified textual query ${q_t}$ and a unified visual query ${q_v}$, each containing the user's essential search demand,  through raw-data level multimodal fusion operations. Subsequently, we seek to optimize an embedding function for the dual query $\left( {q_t}, {q_v} \right)$ and an image embedding function for the target image ${x_t}$, which can ensure the derived representations of the dual query and the target image are as close as possible within the embedding space. This can be formally expressed as follows, 
\begin{equation}
    \mathcal{F}\left( {q_t}, {q_v} | {x_r}, {t_m}\right) \rightarrow \mathcal{H}\left({x_t}\right),\label{eq1}
\end{equation}
where $\mathcal{F}$ and $\mathcal{H}$ denote the \mbox{to-be-learned} functions for embedding the unified dual queries and the target image, respectively.

\subsection{Text-oriented Query Unification}
In this component, we aim to employ an image captioning model to derive a textual description for the reference image.
Then by concatenating the reference image caption and the modification text, the user's multimodal query can be converted into a pure sentence query. 
Specifically, inspired by the remarkable success of the VLP models in image captioning~\cite{xl1,xl2},  we resort to the \mbox{BLIP-2}~\cite{blip2} model to generate a high-quality description for each given reference image\footnote{Any other advanced image captioning model is applicable.}. 
Formally, we have,
\begin{equation}
    t_r = \operatorname{BLIP-2}\left( x_r  \right),
    \label{eq2}
\end{equation}
where $t_r$ denotes the generated caption for the reference image  $x_r$. 

Subsequently, as illustrated in Figure~\ref{fig:model}(a), we derive the unified textual query $q_t$ by concatenating the reference image caption and the modification text through the template ``$t_r${\texttt{, but }}$t_m$''. In this way, the derived textual query contains the essential textual description of the reference image and the complete information of the modification text, and leaves the task of user intention reasoning to the powerful CLIP text encoder. Essentially, the text-oriented unification should benefit the CIR cases with complex modification requests, where the modification text plays the dominant role in delivering the user's search demand.  

\begin{figure}[!t]
	\includegraphics[width=0.93\linewidth]{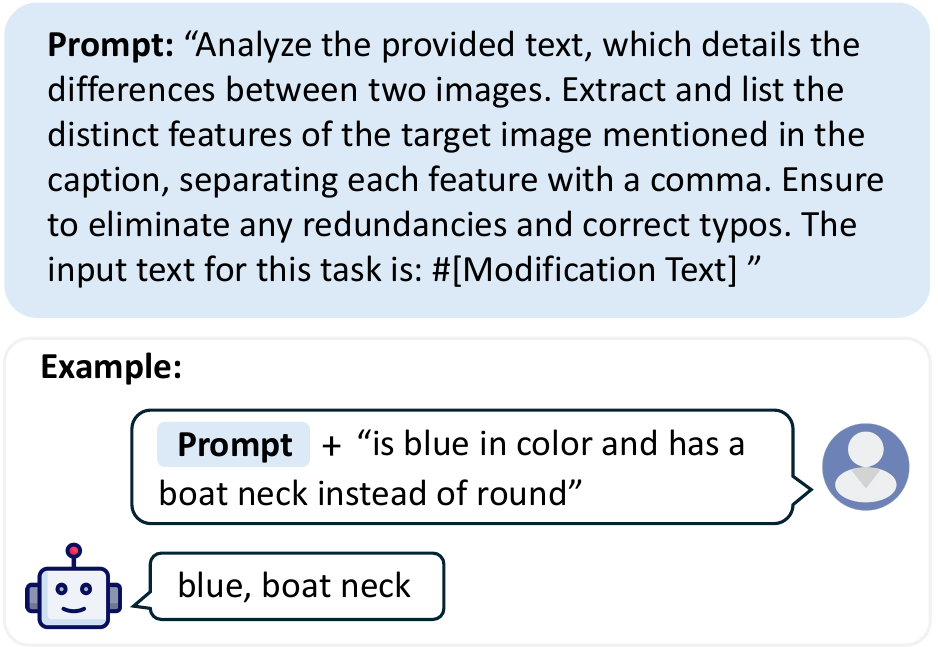}
        \vspace{-0.2em}
	\caption{Illustration of our designed prompt and an example for the key words extraction.}
        \vspace{-1em}
	\label{fig:prompt}
\end{figure}

\subsection{Vision-oriented Query Unification}
In this part, we aim to derive a unified image query that encapsulates the user's search demand based on the input multimodal query. One straightforward solution for deriving the unified image query is to synthesize a new image that depicts the user's overall search demand with the image manipulation technology~\cite{imageedit1, yx}, where the modification text is used to manipulate the reference image. 
However, to ensure the image manipulation effect, the off-the-shelf generative image manipulation models need to be re-trained on the CIR dataset, which would lead to additional computational costs.
Therefore, we ingeniously design a simple but effective vision-oriented query unification method, that is first extracts the key words of the modification text that indicate the desired attributes of the target image, and directly writes them on the reference image. 
This approach ensures that the visual details of the reference image are completely retained, and the crucial target desired attributes mentioned in the modification text are embedded in the image. In fact, it has been proven that CLIP image encoder has great potential in OCR~\cite{clipocr1, clipocr2}. Accordingly, we expect the CLIP image encoder to fulfil the user intention reasoning given the unified image query. 


It is worth mentioning that the modification text usually contains both the target desired attributes as well as the original attributes of the reference image that should be replaced. Therefore, to extract the target image description words from the modification text, it is essential to discern which parts pertain to the reference image and should be discarded, and which parts are relevant to the target image and thus should be extracted. This process involves careful reasoning to accurately separate and identify the relevant segments of the modification text. Inspired by the compelling success of the large language models (LLM) in various natural language reasoning tasks, we resort to the Gemini-pro~\cite{gemini} to automatically extract the target image description words from the modification text. Specifically, we design a prompt~\cite{prompt} to enable LLM to assist us in accomplishing this goal. The designed prompt and an example are shown in Figure~\ref{fig:prompt}.

Thereafter, as depicted in Figure~\ref{fig:model}(b), we directly write the extracted key words describing the target image on the blank part of the reference image with the help of the computer vision library OpenCV~\cite{opencv}. 
We regard the reference image with modification keywords on it as the unified visual query $q_v$. Intuitively, the obtained visual query contains the visual details of the reference image and the key description of the target image, which is friendly to the CIR cases with simple modification requests. 

\subsection{Linear Adaptive Fusion-based Target Retrieval.}
As aforementioned, modification demands in the context of CIR are diverse, including both complex and simple modifications. In cases with complex modification, many visual properties of the reference image need to be altered, thus the modification text becomes crucial. As a result, unifying the multimodal query into a textual query is expected to exhibit superior retrieval performance. In contrast, in cases with simple modifications, only minor aspects of the reference image need to be changed, thus the reference image plays a more dominant role. Accordingly, the unified visual query is more suitable to handle these cases.  
In light of this, we propose to adaptively consider the unified textual query and visual query. Specifically, we first encode the unified textual query $q_t$ and visual query $q_v$ through the CLIP text encoder and image encoder, respectively. Formally, we have,
\begin{equation}
    \left\{\begin{aligned}
    \mathbf{f}_{textual} &= \operatorname{CLIP_{text}} \left( q_t \right), \\
    \mathbf{f}_{visual} &= \operatorname{CLIP_{image}} \left( q_v \right),
    \end{aligned}\right. \label{eq3}
\end{equation}
where $\mathbf{f}_{textual} \in \mathbb{R}^{D}$ and $\mathbf{f}_{visual} \in \mathbb{R}^{D}$ refer to the extracted textual query feature and visual query feature, respectively.

We then adaptively fuse the features of the unified queries $q_t$ and $q_v$ with a linear weighted addition, where the weights for fusion are learnable based on the given context. It is worth mentioning that in the previous query unification, we have fulfilled the multimodal query fusion at the raw-data level, which is distinct from previous studies that conduct the multimodal query fusion at the embedding level.
In this part, the fusion process works on fusing the two kinds of unified queries for adaptively handling CIR cases with various modification demands. The reason for designing the linear fusion operation is to maintain the fused query embedding still residing in the VLP's original embedding space, in which the target image embedding resides, and hence facilitates the final target retrieval.

Specifically, we fuse the features of the two unified queries with the following linear weighted addition, 
\begin{equation}
  \left\{\begin{aligned}
     \mathbf{f}_q& = \lambda * \mathbf{f}_{textual} + \left(1-\lambda\right) * \mathbf{f}_{visual}, \\
    \lambda& = \sigma \left( \operatorname{MLP}\left( \left[ \mathbf{f}_{textual} \| \mathbf{f}_{visual} \right] \right)\right),
    \end{aligned}\right. \label{eq4}
\end{equation}
where  $\mathbf{f}_q \in \mathbb{R}^{D}$ denotes the final composed query feature. $\lambda$ is the trade-off weight for balancing the importance of the unified textual query and unified visual query.
$\left[ \cdot \| \cdot \right]$ denotes the feature concatenation operation, $\operatorname{MLP}$ refers to the multi-layer perceptron, and $\sigma$ represents the Sigmoid activate function.

Finally, the task boils down to pushing the composed query feature close to the corresponding target image feature in the embedding space. Here we resort to the commonly used batch-based classification loss as the optimization function. 
The essence of batch-based classification loss is to closely align the composed query feature with the target image feature within a mini-batch, while simultaneously distancing it from the features of other images. Here we also derive the target image feature through the CLIP image encoder as follows,
\begin{equation} 
    \mathbf{f}_t = \operatorname{CLIP_{image}} \left( x_t \right),
   \label{eq6}
\end{equation}
where $\mathbf{f}_t \in \mathbb{R}^{D}$ is the target image feature. Mathematically, we have the loss function for optimizing the composed query feature and target image feature as follows,
\begin{equation} 
\mathcal{L}=\frac{1}{B} \sum_{i=1}^{B}-\log \left\{\frac{\exp \left\{\operatorname{cos} \left(\mathbf{f}_{qi}, \mathbf{f}_{ti}\right) / \tau \right\}}{\sum_{j=1}^{B} \exp \left\{\operatorname{cos} \left(\mathbf{f}_{qi}, \mathbf{f}_{tj}\right) / \tau \right\}}\right\},
   \label{eq7}
\end{equation}
where the subscript $i$ refers to the $i$-th triplet sample in the mini-batch, $B$ is the batch size, $\operatorname{cos}\left(\cdot, \cdot\right)$ denotes the cosine similarity function, and $\tau$ is the temperature factor. For optimization, we parameterize the final objective function for our DQU-CIR as follows,
\begin{equation}
\mathbf{\Theta^{*}}=
\underset{\mathbf{\Theta}}{\arg \min } \left( \mathcal{L} \right),\label{eq16}
\end{equation}
where $\mathbf{\Theta}$ denotes the \mbox{to-be-learned} parameters of our DQU-CIR.

\section{Experiment}

\begin{table*}[t!]
    \centering
    \caption{Performance comparison on FashionIQ with respect to R@$k$($\%$). Averages and standard deviations are reported from $5$ random seed experiments. The best results are colored in \textcolor{defblue}{\textbf{blue}}, while the second-best results are \underline{underlined}.}
    \vspace{-0.5em}
   
    \begin{tabular}{c|l|cc|cc|cc|cc|c}
    \hline 
    \multirow{2}{*}{Split} & \multirow{2}{*}{Method} & \multicolumn{2}{c|}{Dresses} & \multicolumn{2}{c|}{Shirts} & \multicolumn{2}{c|} {Tops\&Tees} & \multicolumn{2}{c|}{Average} & \multirow{2}{*}{Avg.} \\ \cline{3-10}
    &  &  R@$10$ & R@$50$ & R@$10$ & R@$50$ & R@$10$ & R@$50$ & R@$10$ & R@$50$ & \\
    \hline \hline
    
  \rowcolor{gray!5} \cellcolor{white} \multirow{17}{*}{\rotatebox{90}{{VAL-Split}}} 
     & \multicolumn{10}{c}{\textit{\textcolor{gray}{Traditional Model-Based Methods}}} \\
    & TIRG~\cite{tirg} \footnotesize{\textcolor{gray}{(CVPR'19)}} & $14.87$ & $34.66$ & $18.26$ & $37.89$ & $19.08$ & $39.62$ & $17.40$ & $37.39$ &$12.60$ \\
    
    & VAL~\cite{val} \footnotesize{\textcolor{gray}{(CVPR'20)}} & $21.12$ & $42.19$ & $21.03$ & $43.44$ & $25.64$ & $49.49$ & $22.60$ & $45.04$ & $16.49$ \\
    
    &  CIRPLANT~\cite{cirr} \footnotesize{\textcolor{gray}{(ICCV'21)}} & $17.45$ & $40.41$ & $17.53$ & $38.81$ & $21.64$ & $45.38$ & $18.87$ & $41.53$ & $ 30.20 $ \\




    & CLVC-Net~\cite{clvcnet} \footnotesize{\textcolor{gray}{(SIGIR'21)}} & $29.85$ & $56.47$ &$ 28.75$ & $54.76$ & $33.50$ & $64.00 $& $30.70$ &$ 58.41 $ & $ 44.56 $ \\

    & ARTEMIS~\cite{artemis} \footnotesize{\textcolor{gray}{(ICLR'22)}}  & $27.16$ & $52.40$ & $21.78$ & $43.64$ &$ 29.20 $& $54.83 $& $26.05 $& $50.29 $ & $ 38.17 $ \\

    & EER~\cite{tip22} \footnotesize{\textcolor{gray}{(TIP'22)}} & $30.02$ & $55.44$ & $25.32$ & $49.87$ & $33.20$ & $60.34$ & $29.51$ & $55.22$ & $ 42.37 $ \\


    & CRR~\cite{crr} \footnotesize{\textcolor{gray}{(MM'22)}} & $30.41$ & $57.11$ & $30.73$ & $58.02$ & $33.67$ & $64.48$ & $31.60$ & $59.87$ & $ 45.74 $ \\

    & AMC~\cite{amc} \footnotesize{\textcolor{gray}{(TOMM'23)}} & $31.73$ & $59.25$ & $30.67$ & $59.08$ & $36.21$ & $66.60$ & $32.87$ & $61.64$ & $ 47.26 $ \\

    & CRN~\cite{crn} \footnotesize{\textcolor{gray}{(TIP'23)}} & $32.67$ & $59.30$ & $30.27$ & $56.97$ & $37.74$ & $65.94$ & $33.56$ & $60.74$ & $ 47.15 $ \\   


    & CMAP~\cite{cmap} \footnotesize{\textcolor{gray}{(TOMM'24)}} & $36.44$ & $64.25$ & $34.83$ & $60.06$ & $41.79$ & $69.12$ & $37.64$ & $64.42$ & $ 51.03 $  \\
    \cdashline{2-11} 
    
   \rowcolor{gray!5} \cellcolor{white} & \multicolumn{10}{c}{\textit{\textcolor{gray}{VLP Model-Based Methods}}} \\

    & Prog. Lrn.~\cite{pron} \footnotesize{\textcolor{gray}{(SIGIR'22)}} & $38.18$ & $64.50$ & $48.63$ & $71.54$ & $52.32$ & $76.90$ & $46.37$ & $70.98$ & $ 58.68 $ \\
    
    & TG-CIR~\cite{tgcir} \footnotesize{\textcolor{gray}{(MM'23)}} & $45.22$ & $69.66$ & $52.60$ & $72.52$ & $56.14$ & $77.10$ & $51.32$ & $73.09$ & {$ 62.21 $} \\
    
    & LIMN+~\cite{limn+} \footnotesize{\textcolor{gray}{(TPAMI'24)}} & \underline{$52.11$} & \underline{$75.21$} & \underline{$57.51$} & \underline{$77.92$} & \underline{$62.67$} & \underline{$82.66$} & \underline{$57.43$} & \underline{$78.60$} & \underline{$ 68.02 $} \\


    & SPIRIT~\cite{spirit} \footnotesize{\textcolor{gray}{(TOMM'24)}} & $43.83$ & $68.86$ & $52.50$ & $74.19$ & $56.60$ & $79.25$ & $50.98$ & $74.10$ & {$ 62.54 $} \\

    \rowcolor{gray!15} \cellcolor{white} &  
    \textbf{DQU-CIR} &
    \hspace{-0.4em} \textcolor{defblue}{$\mathbf{57.63} ^ {\pm\scalebox{0.7}{0.24}}$} \hspace{-0.7em} &
    \hspace{-0.4em} \textcolor{defblue}{$\mathbf{78.56} ^ {\pm\scalebox{0.7}{0.50}}$} \hspace{-0.7em} &
    \hspace{-0.4em} \textcolor{defblue}{$\mathbf{62.14} ^ {\pm\scalebox{0.7}{0.66}}$} \hspace{-0.7em} &
    \hspace{-0.4em} \textcolor{defblue}{$\mathbf{80.38} ^ {\pm\scalebox{0.7}{0.15}}$} \hspace{-0.7em} & 
    \hspace{-0.4em} \textcolor{defblue}{$\mathbf{66.15} ^ {\pm\scalebox{0.7}{0.50}}$} \hspace{-0.7em} &
    \hspace{-0.4em} \textcolor{defblue}{$\mathbf{85.73} ^ {\pm\scalebox{0.7}{0.25}}$} \hspace{-0.7em} & 
    \hspace{-0.4em}\textcolor{defblue}{$\mathbf{61.97}^ {\pm\scalebox{0.7}{0.28}}$}\hspace{-1em} &
    \hspace{-0.8em} \textcolor{defblue}{$\mathbf{81.56}^ {\pm\scalebox{0.7}{0.22}}$}\hspace{-0.7em} & 
    \hspace{-0.4em} \textcolor{defblue}{ $\mathbf{71.77}^ {\pm\scalebox{0.7}{0.17}}$} \hspace{-0.7em}
    \\ 
    \hline \hline


    \rowcolor{gray!5} \multirow{11}{*}{\cellcolor{white}\rotatebox{90}{{Original-Split}}}

    & \multicolumn{10}{c}{\textit{\textcolor{gray}{Traditional Model-Based Methods}}}  \\
    & TIRG~\cite{tirg} \footnotesize{\textcolor{gray}{(CVPR'19)}} & $14.13$ & $34.61$ & $13.10$ & $30.91$ & $14.79$ & $34.37$ & $14.01$ & $33.30$ &$23.66$ \\


    & ARTEMIS~\cite{artemis} \footnotesize{\textcolor{gray}{(ICLR'22)}}  & $25.68$ & $51.05$ & $21.57$ & $44.13$ &$ 28.59 $& $ 55.06 $& $ 25.28 $& $50.08 $ & $ 37.68$ \\


     \cdashline{2-11} 

    & \multicolumn{10}{c}{\cellcolor{gray!5}\textit{\textcolor{gray}{VLP Model-Based Methods}}} \\
    & CLIP4CIR~\cite{clip4cir} \footnotesize{\textcolor{gray}{(CVPR'22)}} & $31.63$ & $56.67$ & $36.36$ & $58.00$ & $38.19$ & $62.42$ & $35.39$ & $59.03$ & $47.21$ \\
    
    & Prog. Lrn.~\cite{pron} \footnotesize{\textcolor{gray}{(SIGIR'22)}} & $33.60$ & $58.90$ & $39.45$ & $61.78$ & $43.96$ & $68.33$ & $39.02$ & $63.00$ & $51.01$ \\
    & FAME-ViL~\cite{fame_vil} \footnotesize{\textcolor{gray}{(CVPR'23)}} & \underline{$42.19$} & \underline{$67.38$} & \underline{$47.64$}& \underline{$68.79$} & \underline{$50.69$} & \underline{$73.07$} & \underline{$46.84$} &\underline{$ 69.75 $} & \underline{$ 58.30 $} \\


    & SPIRIT~\cite{spirit} \footnotesize{\textcolor{gray}{(TOMM'24)}} & $39.86$ & $64.30$ & $44.11$ & $65.60$ & $47.68$ & $71.70$ & $43.88$ & $67.20$ & $ 55.54$ \\

    & BLIP4CIR+Bi~\cite{Liu_2024_WACV} \footnotesize{\textcolor{gray}{(WACV'24)}} \hspace{-0.1em}& $42.09$ & $67.33$ & $41.76$ & $64.28$ & $46.61$ & $70.32$ & $43.49$ & $67.31$ & $55.40 $ \\

    \rowcolor{gray!15} \cellcolor{white} &
    \textbf{DQU-CIR} & 
    \hspace{-0.4em}\textcolor{defblue}{$\mathbf{51.90} ^ {\pm\scalebox{0.7}{0.64}}$} \hspace{-0.7em} &
    \hspace{-0.4em}\textcolor{defblue}{$\mathbf{74.37} ^ {\pm\scalebox{0.7}{0.39}}$} \hspace{-0.7em} &
    \hspace{-0.4em}\textcolor{defblue}{$\mathbf{53.57} ^ {\pm\scalebox{0.7}{0.27}}$} \hspace{-0.7em} &
    \hspace{-0.4em}\textcolor{defblue}{$\mathbf{73.21} ^ {\pm\scalebox{0.7}{0.34}}$} \hspace{-0.7em} &
    \hspace{-0.4em}\textcolor{defblue}{$\mathbf{58.48} ^ {\pm\scalebox{0.7}{0.46}}$} \hspace{-0.7em} &
    \hspace{-0.4em}\textcolor{defblue}{$\mathbf{79.23} ^ {\pm\scalebox{0.7}{0.29}}$} \hspace{-0.7em} &
    \hspace{-0.4em}\textcolor{defblue}{$\mathbf{54.65} ^ {\pm\scalebox{0.7}{0.38}}$} \hspace{-1em}&
    \hspace{-0.8em}\textcolor{defblue}{$\mathbf{75.60} ^ {\pm\scalebox{0.7}{0.18}}$} \hspace{-0.7em}&
    \hspace{-0.4em}\textcolor{defblue}{$\mathbf{65.13} ^ {\pm\scalebox{0.7}{0.14}}$}\hspace{-0.7em} \\ 
    \hline
    \end{tabular}
    \vspace{-0.9em}
    \label{tab:exp_fashioniq}
\end{table*}

\subsection{Experimental Settings}
\subsubsection{Datasets.} We chose four public datasets to evaluate our DQU-CIR, including three fashion-domain datasets including FashionIQ~\cite{fashioniq}, Shoes~\cite{shoes}, and Fashion200K~\cite{fashion200k}, as well as an open-domain dataset CIRR~\cite{cirr}. 

\subsubsection{Implementation Details.}
DQU-CIR employs the pre-trained CLIP (ViT-H/$14$ version) for feature extraction and is trained by the AdamW optimizer. For the FashionIQ, Fashion200K, and CIRR datasets, CLIP parameters are fine-tuned at a learning rate of $1e-6$, while other DQU-CIR parameters are optimized at $1e-4$ for effective convergence. For the Shoes dataset, these learning rates are adjusted to $5e-6$ for CLIP and $5e-5$ for other parameters. The feature dimension $D$ is set to $1024$, the batch size $B$ is fixed at $16$, and the temperature factor $\tau$ in Eqn.($\ref{eq7}$) is set to $0.1$ for all four datasets. All experiments are conducted using PyTorch on a server equipped with a single A100-40G GPU. 

\subsubsection{Evaluation.} We adhered to the standard evaluation protocols for each dataset and utilized the Recall@$k$ (R@$k$) metric for a fair comparison.
For the FashionIQ dataset, we followed the challenge's evaluation criteria~\cite{fashioniq}, focusing on R@$10$ and R@$50$ across its three categories. 
Notably, FashionIQ comprises two evaluation protocols: the VAL-Split~\cite{val} and the Original-Split~\cite{fashioniq}. The VAL-Split is introduced by the early-stage CIR study, which builds the candidate image set for testing based on the union of the reference images and target images in all the triplets of the validation set. The Original-Split is recently adopted, which directly uses the original candidate image set provided by the FashionIQ dataset for testing. We reported results for both dataset splits to provide a comprehensive evaluation.
In the case of the Shoes and Fashion200K datasets, aligned with prior studies~\cite{val,clvcnet,artemis,tgcir},  we reported R@$1$, R@$10$, R@$50$, and their calculated averages. 
For CIRR, following the precedent set by ~\cite{cirr,artemis}, we reported R@$k$ for $k=1, 5, 10, 50$, R$_{subset}$@$k$ for $k=1, 2, 3$, and the average of R@$5$ and R$_{subset}$@$1$.

\begin{table}[t!]
    \centering
    \caption{Performance comparison on Shoes regarding R@$k$($\%$). Averages and standard deviations are reported from $5$ random seed experiments. The best results are colored in \textcolor{defblue}{\textbf{blue}}, while the second-best results are \underline{underlined}.
    }
    \vspace{-0.7em}
    \begin{tabular}{l|ccc|c}
    \hline 
    Method & R@$1$ & R@$10$ & R@$50$ & Avg. \\
    \hline \hline 
   \rowcolor{gray!5} \multicolumn{5}{c}{\textit{\textcolor{gray}{Traditional Model-Based Methods}}} \\
    TIRG~\cite{tirg} \footnotesize{\textcolor{gray}{(CVPR'19)}} & $12.60$ & $45.45$ & $69.39$ &$42.48$\\
    VAL~\cite{val} \footnotesize{\textcolor{gray}{(CVPR'20)}} & $16.49$ & $49.12$ & $73.53$ & $46.38$\\
    CLVC-Net~\cite{clvcnet} \footnotesize{\textcolor{gray}{(SIGIR'21)}}\hspace{-0.6em} &$17.64$ & $54.39$ & $79.47$ &$50.50$\\
    ARTEMIS~\cite{artemis} \footnotesize{\textcolor{gray}{(ICLR'22)}}  &$18.72$ & $53.11$ & $79.31$ &$50.38$\\
    EER~\cite{tip22} \footnotesize{\textcolor{gray}{(TIP'22)}} & {$20.05$} & $56.02$ & {$79.94$} & $52.00$\\
    CRR~\cite{crr} \footnotesize{\textcolor{gray}{(MM'22)}} & $18.41$ & $56.38$ & $79.92$ & $51.57$\\
    AMC~\cite{amc} \footnotesize{\textcolor{gray}{(TOMM'23)}} & $19.99$ & {$56.89$} & $79.27$ & {$52.05$}\\
    CRN~\cite{crn} \footnotesize{\textcolor{gray}{(TIP'23)}} & $18.92$ & $54.55$ & $80.04$ & $51.17$\\
    CMAP~\cite{cmap} \footnotesize{\textcolor{gray}{(TOMM'24)}} & $21.48$ & $56.18$ & $81.14$ & $52.93$\\
    \hdashline
   \rowcolor{gray!5} \multicolumn{5}{c}{\textit{\textcolor{gray}{VLP Model-Based Methods}}} \\
    Prog. Lrn.~\cite{pron} \footnotesize{\textcolor{gray}{(SIGIR'22)}}\hspace{-0.6em} & $22.88$ & $58.83$ & $84.16$ & $55.29$\\
    TG-CIR~\cite{tgcir} \footnotesize{\textcolor{gray}{(MM'23)}} & \underline{$25.89$} & {$63.20$} & {$85.07$} & \underline{$58.05$} \\
    LIMN+~\cite{limn+} \footnotesize{\textcolor{gray}{(TPAMI'24)}} & $-$ & \underline{$68.37$} & \underline{$88.07$} & $-$\\
    SPIRIT~\cite{spirit} \footnotesize{\textcolor{gray}{(TOMM'24)}} & $-$ & $56.90$ & $81.49$ & $-$\\

    \rowcolor{gray!15} \textbf{DQU-CIR} & \hspace{-0.6em} \textcolor{defblue}{$\mathbf{31.47} ^ {\pm\scalebox{0.6}{1.31}}$} \hspace{-0.85em} & \hspace{-0.2em}\textcolor{defblue}{$\mathbf{69.19} ^ {\pm\scalebox{0.6}{0.99}}$} \hspace{-0.85em} & \hspace{-0.2em} \textcolor{defblue}{$\mathbf{88.52} ^ {\pm\scalebox{0.6}{0.31}}$} \hspace{-0.8em} & \hspace{-0.4em}\textcolor{defblue}{$\mathbf{63.06} ^ {\pm\scalebox{0.6}{0.69}}$}\hspace{-0.4em}
    \\ 
    \hline
    \end{tabular}
    \label{tab:exp_shoes}
    \vspace{-0.5em}

\end{table}

\begin{table}[t!]
    \centering
    \caption{Performance comparison on Fashion200K regarding R@$k$($\%$). Averages and standard deviations are reported from $5$ random seed experiments. The best results are colored in \textcolor{defblue}{\textbf{blue}}, while the second-best results are \underline{underlined}.
    }
    \vspace{-0.5em}
    \begin{tabular}{l|ccc|c}
    \hline 
    Method & R@$1$ & R@$10$ & R@$50$ & Avg. \\
    \hline \hline 
   \rowcolor{gray!5} \multicolumn{5}{c}{\textit{\textcolor{gray}{Traditional Model-Based Methods}}} \\
    TIRG~\cite{tirg} \footnotesize{\textcolor{gray}{(CVPR'19)}} & $14.1$ & $42.5$ & $63.8$ &$40.1$\\
    VAL~\cite{val} \footnotesize{\textcolor{gray}{(CVPR'20)}} & $22.9$ & $50.8$ & $72.7$ & $48.8$\\
    CLVC-Net~\cite{clvcnet} \footnotesize{\textcolor{gray}{(SIGIR'21)}} &$22.6$ & $53.0$ & $72.2$ &$49.3$\\
    ARTEMIS~\cite{artemis} \footnotesize{\textcolor{gray}{(ICLR'22)}}  &$21.5$ & $51.1$ & $70.5$ &$47.7$\\
    EER~\cite{tip22} \footnotesize{\textcolor{gray}{(TIP'22)}} & $-$ & $55.3$ & {$73.4$} & $-$\\
    CRR~\cite{crr} \footnotesize{\textcolor{gray}{(MM'22)}} & \underline{$24.9$} & $56.4$ & $73.6$ & $51.6$\\
    CRN~\cite{crn} \footnotesize{\textcolor{gray}{(TIP'23)}} & $-$ & $53.5$ & $74.5$ & $-$\\
    CMAP~\cite{cmap} \footnotesize{\textcolor{gray}{(TOMM'24)}} & $24.2$ & $56.9$ & $75.3$ & \underline{$52.1$}\\
    \hdashline
   \rowcolor{gray!5} \multicolumn{5}{c}{\textit{\textcolor{gray}{VLP Model-Based Methods}}} \\
    LIMN~\cite{limn+} \footnotesize{\textcolor{gray}{(TPAMI'24)}} & $-$ & \underline{$57.2$} & \underline{$76.6$} & $-$\\
    SPIRIT~\cite{spirit} \footnotesize{\textcolor{gray}{(TOMM'24)}} & $-$ & $55.2$ & $73.6$ & $-$\\

    \rowcolor{gray!15} \textbf{DQU-CIR} & \textcolor{defblue}{$\mathbf{36.8} ^ {\pm\scalebox{0.6}{3.8}}$} \hspace{-0.89em}  & \textcolor{defblue}{$\mathbf{67.9} ^ {\pm\scalebox{0.6}{2.1}}$} \hspace{-0.85em} & \textcolor{defblue}{$\mathbf{87.8} ^ {\pm\scalebox{0.6}{0.3}}$} \hspace{-0.6em}  & \textcolor{defblue}{$\mathbf{64.1} ^ {\pm\scalebox{0.6}{1.7}}$}\hspace{-0.7em}\\ 
    \hline
    \end{tabular}
    \label{tab:exp_fashion200k}
\vspace{-0.7em}
\end{table}

\begin{table*}[ht!]
    \centering \caption{Performance comparison on CIRR with respect to R@$k$($\%$) and R$_{subset}$@$k$($\%$).The random seed is set to $42$ because excessive submissions were being rejected by the evaluation server during multiple experiments. The best results are colored in \textcolor{defblue}{\textbf{blue}}, while the second-best results are \underline{underlined}.
    }\label{tab:exp_cirr}
\vspace{-0.8em}
    \begin{tabular}{l|cccc|ccc|c}
    \hline 
    \multirow{2}{*}{Method} &\multicolumn{4}{c|}{\textbf{R@$k$}} &\multicolumn{3}{c|}{\textbf{R$_{subset}$@$k$}} & \multirow{2}{*}{(R@$5$ + R$_{subset}$@$1$) / $2$} \\ \cline{2-8}
    & $k=1$ & $k=5$ & $k=10$ & $k=50$ & $k=1$ & $k=2$ & $k=3$ \\
    \hline \hline 
    \rowcolor{gray!5} \multicolumn{9}{c}{\textit{\textcolor{gray}{Traditional Model-Based Methods}}} \\
    TIRG~\cite{tirg} \footnotesize{\textcolor{gray}{(CVPR'19)}} & $14.61$ & $48.37$ & $64.08$ & $90.03$ & $22.67$ & $44.97$ & $65.14$ & $35.52$\\
    CIRPLANT~\cite{cirr} \footnotesize{\textcolor{gray}{(ICCV'21)}}  & $15.18$ & $43.36$ & $60.48$ & $87.64$ & $33.81$ & $56.99$ & $75.40$ & $38.59$\\
    ARTEMIS~\cite{artemis} \footnotesize{\textcolor{gray}{(ICLR'22)}} & $16.96$ & $46.10$ & $61.31$ & $87.73$ &$ 39.99 $& $62.20 $& $75.67 $ & $43.05$\\
    \hdashline
    \rowcolor{gray!5} \multicolumn{9}{c}{\textit{\textcolor{gray}{VLP Model-Based Methods}}} \\
    CLIP4CIR~\cite{clip4cir} \footnotesize{\textcolor{gray}{(CVPR'22)}}  & $33.59$ & $65.35$ & $77.35$ & $95.21$ & $62.39$ & $81.81$ & $92.02$ & $63.87$\\
    TG-CIR~\cite{tgcir} \footnotesize{\textcolor{gray}{(MM'23)}} & \underline{${45.25}$} & {\textcolor{defblue}{$\mathbf{78.29}$}} & \underline{${87.16}$} & \underline{${97.30}$} & ${72.84}$ & ${89.25}$ & ${95.13}$ & \underline{${75.57}$} \\
    LIMN+~\cite{limn+} \footnotesize{\textcolor{gray}{(TPAMI'24)}} & ${43.33}$ & ${75.41}$ & ${85.81}$ & ${97.21}$ & ${69.28}$ & ${86.43}$ & ${94.26}$ & ${72.35}$ \\

    SPIRIT~\cite{spirit} \footnotesize{\textcolor{gray}{(TOMM'24)}} & ${40.23}$ & ${75.10}$ & ${84.16}$ & ${96.88}$ & \underline{${73.74}$} & $\underline{89.60}$ & $\textcolor{defblue}{\mathbf{95.93}}$ & ${74.42}$ \\
    BLIP4CIR+Bi~\cite{Liu_2024_WACV} \footnotesize{\textcolor{gray}{(WACV'24)}} & ${40.15}$ & ${73.08}$ & ${83.88}$ & ${96.27}$ & ${72.10}$ & ${88.27}$ & \textcolor{defblue}{${\mathbf{95.93}}$} & ${72.59}$ \\
    
    \rowcolor{gray!15} \textbf{TQU-CIR} \footnotesize{\textcolor{gray}{(unified textual query only)}} & ${44.05}$ & $75.21 $ & $ 85.01$ & ${96.63 }$ & \textcolor{defblue}{$\mathbf{76.41}$} & \textcolor{defblue}{ $\mathbf{90.53}$} & \underline{$95.90$} & \textcolor{defblue}{$\mathbf{75.81}$} \\
    \rowcolor{gray!15} \textbf{VQU-CIR} \footnotesize{\textcolor{gray}{(unified visual query only)}} & ${32.87}$ & $64.80 $ & $77.06$ & ${95.40}$ & $58.10$ & $78.72$ & $89.76$ & $61.45$ \\
    \rowcolor{gray!15} \textbf{DQU-CIR} & $\textcolor{defblue}{{\mathbf{46.22}}}$ & \underline{$78.17$} & $\textcolor{defblue}{\mathbf{87.64}}$ & \textcolor{defblue}{$\mathbf{97.81}$} & $70.92$ & $87.69$ & $94.68$ & $74.55$ \\
    \hline
    \end{tabular}
\vspace{-0.4em}
\end{table*}

\subsection{Performance Comparison}
We compared DQU-CIR with the following two baseline groups: traditional model-based methods and VLP model-based methods. The first group methods utilize conventional models like ResNet and LSTM for feature extraction, including TIRG~\cite{tirg}, VAL~\cite{val}, CLRPLANT~\cite{cirr}, CLVC-Net~\cite{clvcnet}, ARTEMIS~\cite{artemis}, EER~\cite{tip22}, CRR~\cite{crr}, AMC~\cite{amc}, CRN~\cite{crn}, CMAP~\cite{cmap}. While the second group leverages advanced VLP models, such as CLIP and BLIP,  as the feature extraction backbones, including CLIP4CIR~\cite{clip4cir}, Prog.Lrn.~\cite{pron}, FAME-ViL~\cite{fame_vil}, TG-CIR~\cite{tgcir}, LIMN+~\cite{limn+}, SPIRIT~\cite{spirit}, and BLIP4CIR+Bi~\cite{Liu_2024_WACV}.
Tables~\ref{tab:exp_fashioniq}, \ref{tab:exp_shoes}, \ref{tab:exp_fashion200k}, and \ref{tab:exp_cirr} summarize the performance comparison on the four datasets. Note that some baseline methods only report results for selected datasets in their original papers. Accordingly, results for datasets not covered by the baseline methods are omitted from our comparison.

Our observations are presented as follows. 
1) Traditional model-based methods are generally outperformed by VLP model-based methods across four datasets. This trend underscores the advantages of the VLP model's superior multimodal encoding capabilities. Such capabilities are pivotal for effective multimodal fusion and cross-modal retrieval in CIR tasks. 
2) DQU-CIR consistently surpasses all baseline methods on the three fashion domain datasets by a significant margin. Specifically, DQU-CIR achieves absolute improvements of $5.51\%$, $11.72\%$, $8.63\%$, and $23.0\%$ over the best baseline for the Avg. metric on FashionIQ VAL-Split, FashionIQ Original-Split, Shoes, and Fashion200K, respectively.
These advancements highlight the effectiveness of fusing the multimodal query at raw-data level, and reflects the remarkable natural language reasoning capability of the CLIP text encoder and the impressive OCR capability of the CLIP image encoder.
3) DQU-CIR exhibits comparable performance on the open-domain CIRR dataset, unlike the notable performance improvements observed on fashion-domain datasets. To delve deeper into the underlying reasons, we conducted supplementary experiments with two derivatives of our model: using the unified textual query (TQU-CIR) and using the unified visual query only (VQU-CIR). The results are shown in Table~\ref{tab:exp_cirr}. As can be seen, TQU-CIR outperforms with a large margin compared to VQU-CIR, with $14.36\%$ towards the (R@$5$ + R$_{subset}$@$1$) / $2$ metric. This may be due to the fact that the modification demands in the open-domain CIRR dataset are considerably more complex than those in the fashion-domain datasets. This modification complexity emphasizes the essential role of modification text in target image retrieval, while the reference image's contribution is negligible. Accordingly, the unified visual query proves inadequate in capturing the intricate modification requirements of the CIRR dataset, leading to suboptimal performance. In fact, TQU-CIR even exceeds the full model DQU-CIR in the R$_{subset}$@$k$ metrics, indicating that in cases with extremely complex modification demands, the unified visual query may have a detrimental effect, and relying solely on the unified textual query can yield better outcomes.

\begin{figure}[!t]
	\includegraphics[width=0.95\linewidth]{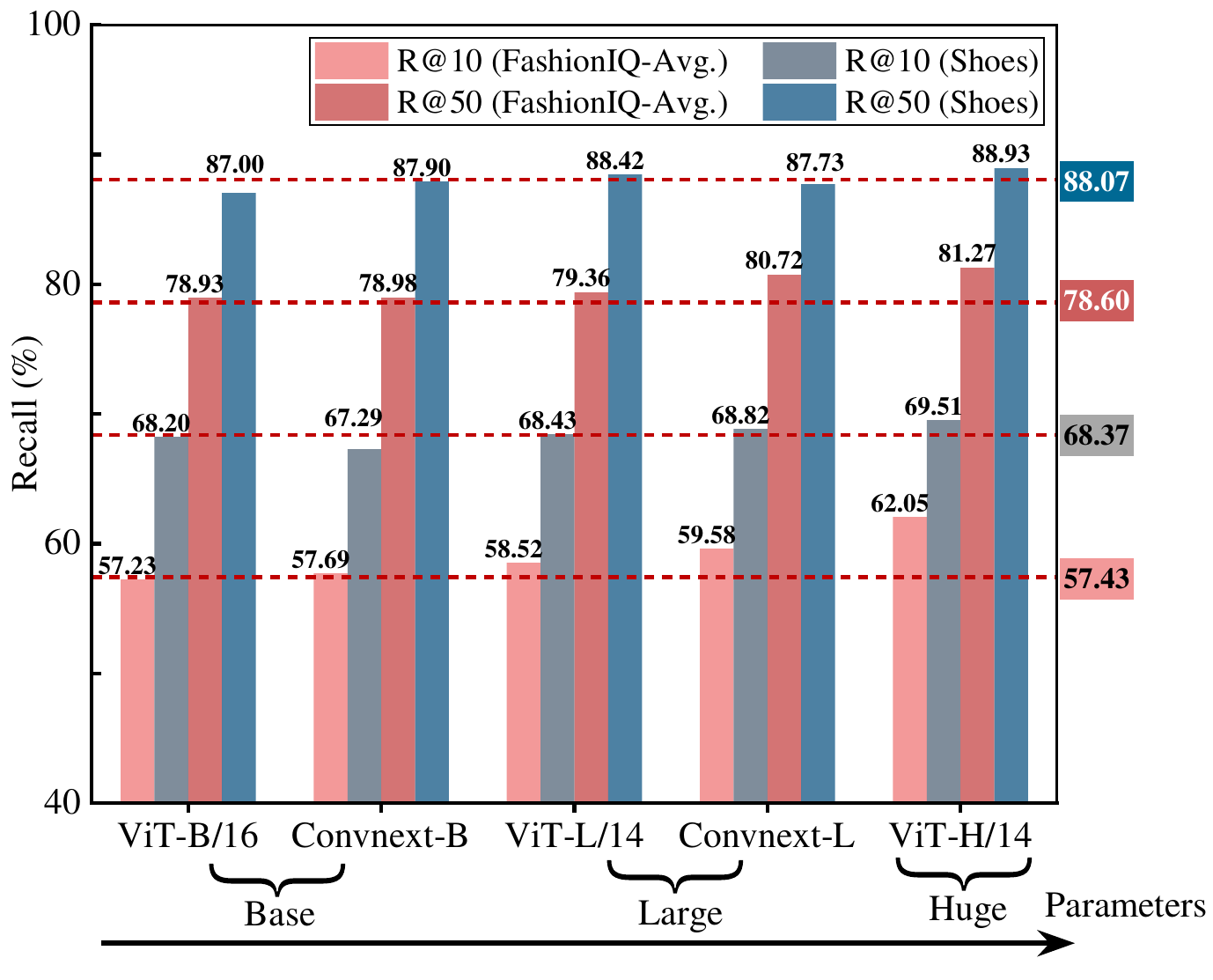}
        \vspace{-0.8em}
	\caption{Performance on FashionIQ-Avg (VAL-Split) and Shoes of DQU-CIR with different versions of CLIP. The horizontal dashed lines denote the best baseline performance. The random seeds are fixed at $42$ across the experiments.}
       \vspace{-1.2em}
	\label{fig:backbone}
\end{figure}

\textbf{On Feature Extraction Backbone.} To provide more insight into the effect of feature extraction backbones, we illustrated the performance of DQU-CIR with different versions of CLIP~\cite{open_clip} on FashionIQ and Shoes in Figure~\ref{fig:backbone}. This includes ViT-based backbones ``ViT-B/16'', ``ViT-L/14'', and ``ViT-H/14'', as well as convolutional neural network (CNN)-based CLIP backbones like ``Convnext-B'' and ``Convnext-L''. As a reference, we also included the results of the best baseline LIMN+~\cite{limn+} for the two datasets, represented by horizontal dashed lines. 
Notably, LIMN+~\cite{limn+}  adopts CLIP ViT-L/14 as the backbone. From Figure~\ref{fig:backbone}, we obtained the following observations. 1) The performance of our model generally improves with larger backbones, this is reasonable as the larger backbones typically have the more powerful multimodal encoding capabilities.
Accordingly, utilizing a larger backbone could potentially enhance the performance. Specifically, we selected the ``ViT-H/14" version as the feature extraction backbone in our experiments. Notably, since we move the multimodal query fusion from the feature level to the raw-data level, which saves computational and memory costs, making it possible for us to deploy the larger backbone under the same computing resources. 
2) With the same ``ViT-L/14'' backbone, our method outperforms the best baseline, LIMN+, which involves complicated feature-level multimodal fusion and sophisticated data augmentation. This reflects the superiority of our well-designed raw-data level multimodal query fusion.

\begin{table*}[t!]
    \centering \caption{Ablation study on FashionIQ and Shoes towards three key components of DQU-CIR. The ``Ref-Img'', ``Mod-Text'', and ``Ref-Cap'' denote the reference image, modification text, and the reference image caption obtained by BLIP-2, respectively. AM\# is the abbreviation of the ablation method number for simplicity. The random seeds are fixed at $42$ across the experiments.}\vspace{-0.6em}
    \label{exp:ablation}
    \begin{tabular}{c|c|clc|cc|cc|cc|cc|c}
    \hline 
   \multirow{3}{*}{AM\#} & \multirow{3}{*}{Comp.}& \multirow{3}{*}{Ref-Img} & \multirow{3}{*}{Mod-Text} & \multirow{3}{*}{Ref-Cap} & \multicolumn{6}{c|}{FashionIQ (VAL-Split)} & \multicolumn{2}{c|}{\multirow{2}{*}{Shoes}} & \multirow{3}{*}{Avg.} \\
    \cline{6-11}
  & & & & & \multicolumn{2}{c|}{Dresses} & \multicolumn{2}{c|}{Shirts} & \multicolumn{2}{c|}{Tops\&Tees} &  & &   \\ 
    \cline{6-13} & & & & & R@$10$ & R@$50$ & R@$10$  & R@$50$ & R@$10$  & R@$50$  & R@$10$  & R@$50$  &\\
    \hline \hline

  $1$ & \multirow{3}{*}{(a)} & & \Checkmark &    & $41.70$ & $66.04$ & $44.95$ & $65.85$ & $51.81$ & $75.42$  & $31.01$ & $55.25$ & $54.00$ \\
  $2$ & & & \Checkmark &    \Checkmark  & $45.91$ & $70.30$ & $56.77$ & $77.33$ & $58.18$ & $81.64$ & $50.54$ & $78.82$ & $64.94$\\
  $3$ &  & &  &    \Checkmark  & $3.47$ & $10.36$ & $15.90$ & $29.20$ & $9.38$ & $20.45$ & $10.51$ & $30.38$ & $16.21$ \\
    \hdashline
  $4$ &  \multirow{5}{*}{(b)} & \Checkmark & & & $12.89$ & $28.01$ & $23.45$ & $40.19$ & $18.71$ & $35.14$ &  $37.37$ & $62.52$ & $32.29$ \\
  $5$ & & \Checkmark & \Checkmark \small{{(Blue)}} &  & $ {43.83} $ & ${68.86}$ & $ {51.91}$ & ${ 72.37}$  & $ {55.43 }$ & ${77.00}$ & ${66.38}$ & ${85.69}$ & $ {65.18} $ \\
  $6$ & & \Checkmark & \Checkmark \small{(Green)} &  &$44.42 $ &$68.82 $ & $ 51.96$& $ 72.18$ & $ 54.26 $ & $ 76.70$ &$ 65.59$ & $85.75 $ & $64.96$\\
  $7$ & & \Checkmark & \Checkmark \small{(Red)} &   & $44.22$ & $67.92$ & $52.31$ & $72.42$ & $53.90 $ & $ 77.00$  & $66.33$ & $ 84.89 $ & $64.87$ \\
  $8$ & & \Checkmark & \Checkmark \small{(Black)} &   & $ 44.52 $ & $68.42$ & $51.42$ & $ 72.87$  & $54.26$ & $76.95$ & $65.81$ & $85.29$ & $64.94$ \\

   \hdashline
   $9$ & (c) & \multicolumn{3}{c|}{Average addition} & $54.98$ & $77.59$ &$61.24$&$80.23$&$64.97$&$84.80$ &$68.60$&$88.07$&$72.56$\\
    \hline
    \rowcolor{gray!15} \multicolumn{5}{c|}{\textbf{DQU-CIR}}  & $\mathbf{56.67}$ & $\mathbf{78.14}$ & $\mathbf{62.46}$ & $\mathbf{80.42}$ &$\mathbf{67.01}$&$\mathbf{85.26}$&  $\mathbf{69.51}$&$\mathbf{88.93}$&$\mathbf{73.55}$ \\
    \hline
    \end{tabular}\vspace{-0.8em}
\end{table*}

\subsection{Ablation Study}
To verify the influence of each component in our model, we conducted ablation study experiments on FashionIQ and Shoes, as shown in Table~\ref{exp:ablation}. Specifically, we compared DQU-CIR with the following ablation methods on three components.
\begin{itemize}[leftmargin=10pt]
    \item \textbf{On component (a) :} In the text-oriented unification component, we concatenated the generated reference image caption and the modification text as the unified text query.  To gain deep insights of this component, we conducted the target image retrieval with three types of queries, including modification text only (\textbf{AM1}), unified textual query only (\textbf{AM2}), and reference image caption only (\textbf{AM3}).
    \item \textbf{On component (b) :} In the vision-oriented unification component, we directly write the target image description on the reference image. 
    To investigate this component, we conducted the target image retrieval with the reference image only (\textbf{AM4}) as a base comparison. In addition, we further investigated the influence of different font colors utilized for writing on the reference image (\textbf{AM5-AM8}).
    \item \textbf{On component (c) :} In the linear adaptive fusion-based target retrieval component, we targeted at validating the effectiveness of the linear weighted addition function. Specifically, we employed the average addition for fusing the two unified queries by setting $\lambda=0.5$ in Eqn.($\ref{eq4}$) (\textbf{AM9}). 
\end{itemize}

From the ablation study results listed in Table~\ref{exp:ablation}, we gained the following observations. 
1) \textbf{AM2} performs much better than both \textbf{AM1} and \textbf{AM3}. This is reasonable as both the reference image and the modification text convey partial user search intention.  
2) \textbf{AM4} shows much inferior results than \textbf{AM5-AM8}. The significant performance gap demonstrates the effectiveness of directly writing target image description words on the reference image to fuse the multimodal query. This also reflects the potential OCR ability of the CLIP image encoder.  
3) \textbf{AM5-AM8} achieve comparable performance, which implies that the font color does not significantly influence the vision-oriented query unification performance. Notably, in all the other experiments, we adopted the blue font, which achieves slightly higher performance.
4) \textbf{AM9} performs consistently worse than the full model.  This proves that the designed linear weighted addition can adaptively fuse the unified textual query and unified visual query for handling cases with various modification demands. 
5) In fact, \textbf{AM2} and \textbf{AM5} correspond to the two derivatives TQU-CIR and VQU-CIR, which only use the text-oriented unified query and the vision-oriented unified query, respectively. We noticed that \textbf{AM2} and \textbf{AM5} perform closely on the FashionIQ and Shoes dataset, unlike that on the CIRR dataset, where only using the unified textual query performs significantly better than solely using the unified visual query. This suggests that the modification demands in these two datasets are moderate, and the retrieval demands are not highly dependent on the modification text as in the CIRR dataset. Meanwhile, this also confirms the result that our linear weighted addition strategy performs well in balancing the unified textual query and unified visual query to address various modification cases.

\begin{figure*}[t!]
    \centering
	\includegraphics[width=0.95\linewidth]{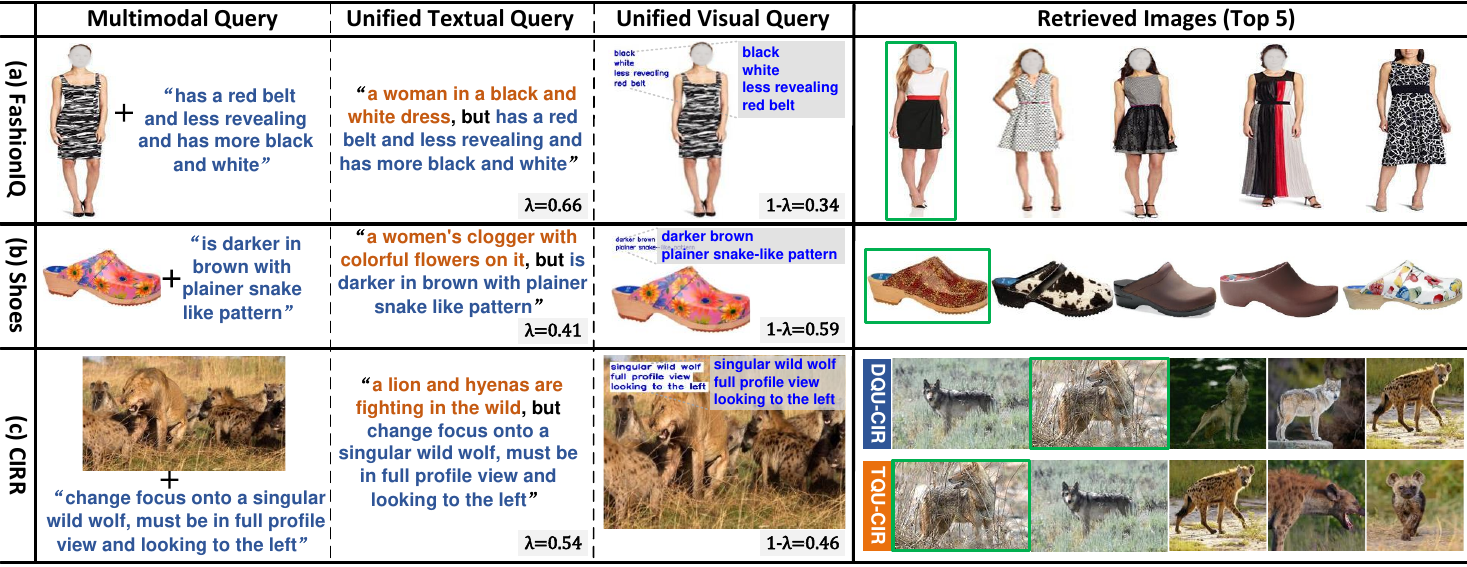}
        \vspace{-0.5em}
	\caption{Case study on (a) FashionIQ, (b) Shoes, and (c) CIRR datasets.}
        \vspace{-1.2em}
	\label{fig:case_study}
\end{figure*}

\subsection{Case Study}

Figure~\ref{fig:case_study} illustrates three examples by DQU-CIR across three datasets. The top $5$ retrieved images are listed, where the green boxes denote the target images. To provide a more intuitive demonstration of our method, we also included the unified textual and visual queries. Specifically, for each unified textual query, the generated reference image caption part is in orange, while the modification text part is in blue. As the key words written on the reference image are too blurry in the top-left of the unified visual query due to limited space, we enlarged them on the top-right for better readability. As can be seen, the reference image captions generated by \mbox{BLIP-2} and the key modification words extracted with \mbox{Gemini-pro} are generally accurate and reasonable.
 The numbers in the bottom-right denote the linear fusion weights. 
As shown in Figure~\ref{fig:case_study}(a), DQU-CIR successfully ranks the target image first given a dress modified in terms of belt, style, and color. The weights show that the unified textual query contributes more than the visual query. This is reasonable as this modification is a little bit complex, and the unified textual query can deliver full modification details by maintaining the original modification text. In Figure~\ref{fig:case_study}(b), our method also places the target image at the top. While in this case, the unified visual query plays a slightly more important role. This is also reasonable since the modification involves simpler changes, just the color and pattern. 
Accordingly, the reference image becomes crucial, and the visual query, which retains all the visual details of the reference image and recognizes the essential modification words, is assigned a higher weight.
Figure~\ref{fig:case_study}(c) shows a failure case.  It can be found that the extensive modification demands render the reference image less influence for the target image retrieval.  To gain more insights, we provided the retrieval results of our derivative TQU-CIR that only utilizes the unified textual query for the target image retrieval. In this case, TQU-CIR successfully ranks the target image at the first place. This may be due to the fact that it better understands the modification demand ``looking to the left''. This suggests that in cases where the search demands are too unique that they can  be directly delivered by the modification text, utilizing only the single unified textual query can achieve promising results. Overall, these observations validate the effectiveness of DQU-CIR in moving the multimodal fusion from the feature level to the raw-data level in the context of CIR.

\section{Conclusion}

In this work, we present a novel dual query unification-based composed image retrieval framework to address the challenging CIR task, which shifts the multimodal fusion from the conventional feature level to the raw-data level.
In particular, we unified the multimodal query by concatenating the reference image caption with the modification text to derive the textual query, and directly writing the key modification words on the reference image to derive the visual query. The derived unified dual queries can be processed by the VLP model's dual encoders, respectively, fully leveraging their multimodal encoding capabilities. Furthermore, the unified textual query excels at handling complex modifications as it preserves the entire modification text. The unified visual query is effective for tackling simple modifications, as it contains the essential visual details of the reference image. To cater to diverse modification demands, we employed a linear weighted addition to combine the dual queries while ensuring that the fused features remain in the original embedding space.
Extensive experiments have been conducted on four public datasets, and the results demonstrate the effectiveness of our method. In the future, we plan to explore more application scenarios with the raw data-level multimodal fusion strategy.

\section*{Acknowledgments}
This work is supported in part by Shenzhen College Stability Support Plan (No.:GXWD20220817144428005), Natural Science Foundation of China (No.:62376137), and Shandong Provincial Natural Science Foundation (No.:ZR2022YQ59). Haokun Wen and Xiaolin Chen acknowledge the support from the China Scholarship Council for pursuing their joint Ph.D. studies at NUS.

\clearpage
\balance

\bibliographystyle{ACM-Reference-Format}
\bibliography{reference}

\end{document}